\documentclass{aa}
\usepackage{astron}
\usepackage{psfig}
\begin{document}

\def\apj{ApJ}
\def\apjl{ApJL}
\def\aap{A\&A}
\def\mnras{MNRAS}
\def\nat{Nature}
\def\aj{AJ}

\thesaurus{ 
3(02.01.1;        
02.19.1;        
11.09.3;        
11.01.2;        
11.03.1;        
13.18.2)        
}
\date{\today}
\title{Reviving Fossil Radio Plasma in Clusters of Galaxies by
Adiabatic Compression in Environmental Shock Waves}
\titlerunning{Reviving Fossil Radio Plasma in Clusters of Galaxies}
\author{Torsten A.  En{\ss}lin\inst{1}, Gopal-Krishna\inst{2}}
\authorrunning{T.A.  En{\ss}lin \& G.-Krishna}
\institute{\inst{1}Max-Planck-Institut f\"{u}r Astrophysik,
Karl-Schwarzschild-Str.1, 85740 Garching, Germany\\ \inst{2}National
Centre for Radio Astrophysics, Tata Institute of Fundamental Research,
Pune University Campus, Ganeshkhind, Pune 411007, India } \maketitle

\begin{abstract}
We give for a plasma with a history of several expansion and
contraction phases an analytical model of the evolution of a contained
relativistic electron population under synchrotron, inverse Compton
and adiabatic energy losses or gains. This is applied to different
scenarios for evolution of radio plasma inside the cocoons of radio
galaxies, after the activity of the central engine has ceased. It is
demonstrated that fossil radio plasma with an age of even up to 2 Gyr
can be revived by compression in a shock wave of large-scale structure
formation, caused during the merging events of galaxy clusters, or by
the accretion onto galaxy clusters.  We argue, that this is a highly
plausible explanation for the observed cluster radio relics, which are
the regions of diffuse radio emission found in clusters of galaxies,
without any likely parent radio galaxy seen nearby. An implication of
this model is the existence of a population of diffuse, ultra-steep
spectrum, very low frequency radio sources located inside and possibly
outside of clusters of galaxies, tracing the revival of aged fossil
radio plasma by the shock waves associated with large-scale structure
formation.
\end{abstract}

\section{Introduction}
The radio cocoons of radio galaxies become rapidly undetectable after
the central engine of the active galactic nucleus ceases to inject
fresh radio plasma, or even earlier \cite[and references
therein]{2000MNRAS.315..381K}.  Radio plasma consists of relativistic electrons
and magnetic fields, which are responsible for the synchrotron radio
emission, and possibly relativistic protons and/or a non-relativistic
thermal gas component. 

Although undetectable, aged radio plasma should still be an
important component of the inter-galactic medium (IGM). We investigate
the possibility of reviving patches of such fossil radio plasma (also
called {\it radio ghosts}, En{\ss}lin \cite*{Ringberg99}) by
compression in a shock wave produced in the IGM by the flows of
cosmological large-scale structure formation. Such shock waves
re-energize the electron population in the fossil radio cocoons,
which can lead to observable synchrotron emission. This is probably
the mechanism responsible for the, so called, {\it cluster radio
relics}, which are patches of diffuse, sometimes polarized radio
emission typically found at peripheral locations in clusters of
galaxies. These cluster radio relics can not be simply relic radio
galaxies, as their name suggests.  The spectral ages of the electron
population are {usually} too short to admit even the nearest galaxy
to have been the parent radio galaxy, which has moved to its present
location with a velocity typical for cluster galaxies. Therefore, a
recent enhancement of the nonthermal radio output of the cluster relic
sources is mandatory.

For reviews on diffuse cluster radio emission (cluster radio relic and
halos, which are believed to be distinct phenomena), see Jaffe
\cite*{1992cscg.conf..109J}, Feretti \& Giovannini
\cite*{1996IAUS..175..333F}, En{\ss}lin et
al. \cite*{1998AA...332..395E}, Feretti \cite*{Feretti.Pune99},
En{\ss}lin \cite*{Pune99}, and Giovannini et
al. \cite*{1999NewA....4..141G}.

\begin{figure}[t]
\begin{center}
\psfig{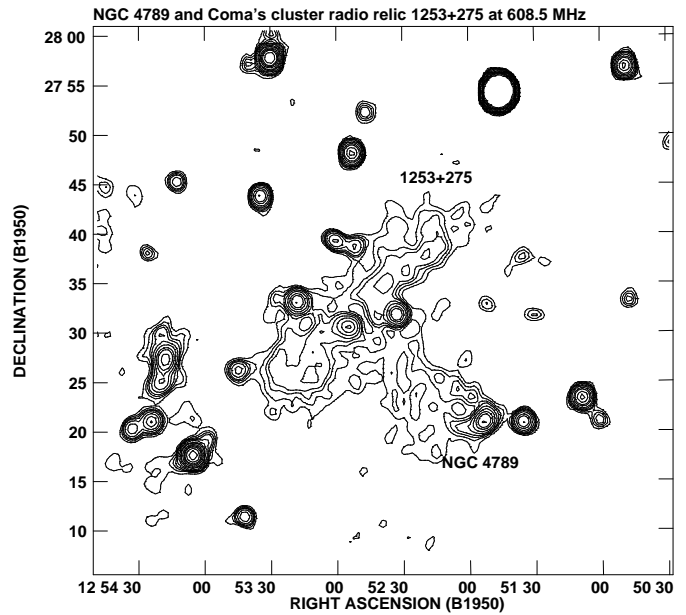}
\end{center}
\caption[]{\label{fig:ngc4789} Radio map of the cluster radio relic
1253+275 in the Coma cluster from Giovannini et
al. \cite*{1991A&A...252..528G} and its probable source of radio
plasma: the narrow angle tailed radio galaxy NGC 4789. The radio tails
of NGC 4789 indicate the direction from which this galaxy came: the
cluster center which is located outside this map. A radio polarization
of 27 \% was detected from the radio relic \cite{1991A&A...252..528G},
but not from the tails of NGC 4789 feeding the relic with radio
plasma. This is a clear signature of the alignment of magnetic
fields enhanced in the shock compression. The polarization vectors are
consistent with a shock wave oriented parallel to the main axis of the
radio relic \cite{1998AA...332..395E}.}
\end{figure}

The connection between the presence of a shock wave and the appearance
of the cluster radio relic phenomena was assumed to be due to Fermi-I
shock acceleration of electrons, in En{\ss}lin et
al. \cite*{1998AA...332..395E} and Roettiger et
al. \cite*{1999ApJ...518..603R}. While this process might work within 
the normal IGM, several arguments favor the possibility that indeed
old fossil radio plasma is revived in the case of cluster relic sources:
\begin{itemize}
\item
Cluster radio relics are extremely rare, whereas shock waves should be
very common within clusters of galaxies. The {dual} requirement of
a shock wave and fossil radio plasma, {for producing} a
cluster radio relic, {would} be an attractive explanation for the rareness of 
the relics.
\item
Fossil radio plasma with existing relativistic electron population and
fairly strong magnetic field appears to have ideal properties to be
brightened up during the shock's passage.
\item
The cluster radio relic 1253+375 near the Coma cluster of galaxies
appears to be fed with radio plasma by the nearby galaxy NGC 4789 (see
Fig \ref{fig:ngc4789} and scenario C in Sec. \ref{sec:scenarios}).
\end{itemize}

But, if indeed the fossil radio plasma and not the normal IGM were to
become radio luminous at a shock wave, the expected very high
sound velocity of that relativistic plasma should forbid the shock in
the ambient medium to penetrate into the radio plasma. Thus, shock
acceleration is not expected to occur there. Instead, the fossil radio
plasma would get adiabatically compressed, and the energy gain of the
electrons is expected to be mainly due to adiabatic heating. It is the
purpose of this work to demonstrate that this process is sufficient to
account for the cluster radio relics.

{Sec. \ref{sec:shocks} and \ref{sec:fossil} discuss the expected
properties of the components of this model: shock waves and fossil
radio plasma.}  In Sec. \ref{sec:formalism} we give a mathematical
formalism to describe the evolution of the radiation spectrum during
the history of a radio cocoon. The different phases of this history
are described in Sec. \ref{sec:phases} and applied to three different
scenarios in Sec. \ref{sec:scenarios}. The discussion of the merits
and potential deficiencies of this model can be found in
Sec. \ref{sec:discussion}.

\section{Structure Formation Shock Waves\label{sec:shocks}}
The flows of the cosmological large-scale structure
formation are predicted to produce frequently shock waves at the
boundaries of clusters and filaments of galaxies and during cluster
merger \cite{1998ApJ...502..518Q,2000ApJ...542..608M}. But only in a few
cases cluster merger shock waves could be detected directly as
temperature structures in the X-ray emitting cluster gas. The best
examples are in Abell 2256 \cite{1994Natur.372..439B} and in Abell
1367 \cite{1998ApJ...500..138D}. The reason for the small X-ray
detectability of shock waves is the necessary simultaneous spatial and
spectral resolution of the instrument, combined with a sufficiently
large collecting area to yield adequate photon statistics from the
often peripheral shock waves in clusters. For instance one of the
best studied cluster, the Coma cluster, is proposed to contain a shock
wave at the location of the diffuse radio source 1253+275 (see
Fig. \ref{fig:ngc4789}; En{\ss}lin et al. \cite*{1998AA...332..395E};
this work). The available X-ray map \cite{1993MNRAS.261L...8W} shows
an extension of the X-ray emitting gas in the direction of this radio
source, but the sensitivity is too low to reveal any structure at its
location. The feasibility of shock wave detections with new
instruments as the XMM experiment is estimated in Tozzi et
al. \cite*{2000ApJ...542..106T}. It is interesting to note that at the locations
of the two above mentioned X-ray detections of shock waves, cluster
radio relics are observed \cite{1998AA...332..395E}. This indicates
that these sources are indeed tracers of shock waves. The importance
of a detailed understanding of these tracers of shock waves is
underscored by the shock wave's possible roles in intergalactic
magnetic field generation
\cite{1997ApJ...480..481K,1998A&A...335...19R}, cosmic ray
acceleration
\cite{1995ApJ...454...60N,1996ApJ...456..422K,1997MNRAS.286..257K,2000A&A...355...51S,Miniatietal2000b},
gamma ray background production \cite{2000Natur.405..156L}, and in probing the
entropic history of the IGM \cite{2000ApJ...542..106T}.

\section{Fossil Radio Plasma\label{sec:fossil}}
In this work it is assumed that the radio plasma in the lobes of radio galaxies
remains mostly intact after release, forming the proposed radio
fossils or ghosts.  This is specified and discussed in the following.
Important for the presented model is that the magnetic fields and
ultra-relativistic electron population continue to remain within the
lobe for up to 2 Gyr. Transport or decay processes of
magnetic fields, and relativistic particles populations need to be
slower than this time scale.

Clearly, the light radio plasma is subject to buoyancy-induced motions
in the cluster gravitational potential. {The action of buoyancy is
e.g. able to explain the morphology of the eastern radio lobe of M87
\cite{churazov200sub}.}  But since this does not separate the
electrons and fields, the fossils would not get
disintegrated. Nevertheless, relative motion at a plasma-field
interface (e.g. the surface of a buoyant fossil) can produce enhanced
mixing \cite{1967ApJ...150..731L} due to finite gyro-radius effects of
the ions in the embedding gas. This should produce a mixing layer of
the order of several gyro-radii of the ions
($10^{-13}...10^{-11}\,$kpc). But this would soften the magnetic
boundary and therefore inhibit the mixing
\cite{1967ApJ...150..731L}. The entrained gas in that boundary layer
might be able to propagate into the fossil following the field lines,
but the large sizes of the fossils ($10...10^3\,$kpc) would forbid
their disintegration due to the gas leakage within the considered time
scale.  The relativistic particles contained in the old radio plasma
have much larger gyro-radii ($10^{-9}\,{\rm kpc}\,(E/{\rm
GeV})\,(B/\mu{\rm G})^{-1}$) than the external gas
particles. Scattering at magnetic irregularities allows the
gyro-center of the particle motion to be displaced, leading to cross
field diffusion required for particle escape. But even if the combined
effect of cross field diffusion and rapid parallel diffusion along
disordered field lines is considered (anomalous diffusion), only very
special turbulent conditions allow an efficient escape of relativistic
particles from radio plasma \cite{Ringberg99}. Strong turbulence
itself is probably more destructive by shredding the magnetized plasma
into smaller pieces. But it has been argued
\cite{Ringberg99,ensslin2000a} that this should stop on scales where
the energy density of the turbulent energy cascade is insufficient to
overcome magnetic forces.

From purely an observational viewpoint there already exists good
evidence that clouds of synchrotron plasma can survive and retain
their identity within the ICM for {\it at least} up to {$\sim 10^8$}
yr, in the form of relic radio sources showing a relaxed structure but
well defined boundaries.  These fairly robust age estimates are based
on interpreting the observed radio spectral steepening in terms of the
inverse Compton losses of the relativisitic electrons against the
ubiquitous cosmic microwave background photons (e.g., Komissarov \&
Gubanov \cite*{1994A&A...285...27K}; Venturi et
al. \cite*{1998MNRAS.298.1113V}; Slee \& Roy
\cite*{1998MNRAS.297L..86S}).

Finally, we note that other energy loss mechanisms of the electrons,
as Coulomb and Bremsstrahlung losses, are inefficient at the
considered energies (1-10 GeV), even if a dense gas component were to
exist inside the radio plasma \cite{ensslin2000a}. {If indeed such
a cold gas component were present inside the radio lobe, Parker-type
instabilities \cite{1966ApJ...145..811P} might occur, helping to shred
the lobe into smaller pieces.  Also Rayleigh-Taylor-type instabilities
of the surface of the light radio plasma might occur, if not supressed
by the tension of the magnetic field lines. But it is questionable if
these instabilities allow a microscopic mixing of the environmental
gas with the radio plasma, since both phases are still separated by
magnetic fields. Further, there is observational evidence that the
density of entrained gas cannot be high in radio lobes and in cluster
radio relics. Many lobes and relics exhibit radio polarization, which
has sometimes values close to the upper limit allowed by synchrotron
theory \cite[e.g.]{1993AJ....105..769H,1996A&A...306..708S}.  No
intrinsic Faraday rotation, or intrinsic Faraday depolarization is
detected in these systems. In the cases of detected depolarization it
seems to be consisten with beam depolarization, or the
Laing-Garrington effect \cite{1988Natur.331..147G}, which is external
to the source. This argues against a dense thermal plasma component.
Nevertheless, we discuss briefly in Sect. \ref{sec:discussion} some
possible implications of strong microscopic mixing for the proposed
model.}

\section{The Formalism\label{sec:formalism}}

The dimensionless momentum of an ultra-relativistic electron $p = P_{\rm e} /(m_{\rm e}
c)$ within the radio plasma changes due to synchrotron losses
proportional to the magnetic energy density $u_B$, inverse Compton
(IC) losses proportional to the cosmic microwave background radiation
field energy density $u_C$, and adiabatic losses or gains connected to
the change of the volume $V$ of the radio plasma:
\begin{equation}
\label{eq:ploss}
- \frac{dp}{dt} = a_{0} \, (u_B + u_C) \,p^2+ 
\frac{1}{3}\,
\frac{1}{V} \,\frac{dV}{dt}\, p \,,
\end{equation}
where $a_{0} = \frac{4}{3}\, \sigma_{\rm T}/(m_{\rm e}\, c)$. We do not
consider bremsstrahlung and Coulomb losses in view of the very low
particle density within the radio plasma. We further assume sufficient
pitch angle scattering to keep the electron pitch angle
distribution isotropic \cite{1973A&A....26..423J}.
After introduction of the compression ratio
\begin{equation}
{C}(t) = V_{0}/V(t) 
\end{equation}
and a temporary change to the variable $\tilde{p} (t) =
{C}(t)^{-1/3}\,p(t) $ Eq.  \ref{eq:ploss} is easily integrable, yielding
finally
\begin{equation}
p(p_0, t) = \frac{p_{0}}{{C}(t)^{-\frac{1}{3}} + p_{0} / p_*(t)}\,.
\end{equation}
Here, we defined the characteristic momentum $p_*(t)$, which is
given by
\begin{equation}
\label{eq:mysol}
\frac{1}{p_*(t)} = a_{0} \, \int_{t_{0}}^t \!\!\!\! dt'
\,(u_B(t')+u_C(t'))\, \left( \frac{{C}(t')}{{C}(t)}
\right)^{\frac{1}{3}}\,.
\end{equation}

If the change in volume can be approximated to be a power-law in time,
\begin{equation}
V(t) = V_{0} \,\left(\frac{t}{t_{0}} \right)^{b} \,,{\rm \,or}\,
{C}(t) = \left(\frac{t}{t_{0}} \right)^{-b}\,,
\end{equation}
which we will assume in the following, an analytic solution to
Eq. \ref{eq:ploss} was already given in Kaiser et
al. \cite*{1997MNRAS.292..723K}.  We assume further the photon energy
density to be constant, which is reasonable in our application since
it is dominated by the cosmic microwave background, which does not
change significantly on the time-scales considered here. Further, we
assume that the magnetic field energy density scales as
\begin{equation}
u_B(t) = u_{B,{0}} \, (V/V_{0})^{-4/3} = u_{B,{0}} \, (t/t_{0})^{- 4\,b /3} \,,
\end{equation}
as one would expects for an isotropic expansion of the magnetized
plasma. These assumptions allow one to evaluate the integral in
Eq. \ref{eq:mysol}. We exclude the cases $b = 3$ and $b = 3/5$
for simplicity only (they produce logarithms instead of power-law relations),
and write the characteristic electron momentum in a compact form:
\begin{equation}
p_*(t) = \frac{{C}^{\frac{1}{3}}}{a_{0} \, t \, \left(
\frac{{C}^{5/3} - {C}^{1/b}}{1- 5\,b /3} \,u_{B{0}} + \frac{{C}^{1/3} -
{C}^{1/b}}{1- b /3}\, u_{C} \right)}
\end{equation}
(${C} = {C}(t)$ for brevity). It is obvious that the synchrotron- and IC-
cooling produces a sharp upper cutoff in the electron distribution
$f(p,t)$ at $p_*(t)$ even if the original distribution $f_{0}(p_0)$ at
time $t_{0}$ extended to infinity. The electron density
per volume and momentum $f(p,t)\, dp\,dV$ for $p<p_*(t)$ is given by
\begin{equation}
\label{eq:spec1}
f(p,t) = f_{0}(p_{0}(p,t)) \frac{\partial p_{0}(p,t)}{\partial
p}\,{C}(t)\,,
\end{equation}
where 
\begin{equation}
p_{0}(p,t) = \frac{ p\,{C}(t)^{-\frac{1}{3}}}{1 - p/p_*(t)}\,. 
\end{equation}
If the original distribution function was a power-law
\begin{equation}
f_{0}(p_{0}) = \tilde{f}_{0} \, p_0^{-\alpha_{\rm e}}
\end{equation}
for $p_{\rm min\,0} < p_0 < p_{\rm max\, 0}$ the resulting spectrum is
\begin{equation}
\label{eq:spec2}
f(p,t) = \tilde{f}_{0}  \, {C}(t)^{\frac{\alpha_{\rm e}
+2}{3}}\, p^{-\alpha_{\rm e}}\, \left( 1 - p/p_{*}(t) \right)^{\alpha_{\rm e} -2}
\end{equation}
for $p_{\rm min}(t)= p(p_{\rm min\,0},t) < p < p_{\rm max}(t) =
p(p_{\rm max\, 0},t)$.

We are interested in the situation where several phases of cooling
characterized by different expansion or contraction rates and
durations shaped the electron distribution. We write $p_1 = p(t_1)$
for the momentum of an electron originally at $p_{0}$ after phase 1,
which is characterized by the compression during this phase ${C}_{0\,1}
= {C}(t_1)$ and $p_{*0\,1} = p_*(t_1)$, $p_2$ for the momentum of the
same electron after phase 2, characterized by ${C}_{1\,2}$ and
$p_{*1\,2}$, and so on. It is straightforward to show that the final
electron momentum $p_n$ after $n$ such phases can still be written in
the form
\begin{equation}
p_n(p_0) = \frac{p_{0}}{({C}_{0\,n})^{-\frac{1}{3}} + p_{0} / p_{*0\,n}}\,,
\end{equation}
where 
\begin{equation}
{C}_{0\, n} = \prod_{i=1}^{n}\, {C}_{i-1\, i} = \prod_{i=1}^{n}\,
\frac{V_i-1}{V_{i}} = \frac{V_{0}}{V_n}
\end{equation}
is the compression ratio between original and final configuration and
maximal final momentum is given by
\begin{equation}
\frac{1}{p_{*0\,n}} = \sum_{i=1}^{n}\, \frac{({C}_{0\,
i-1})^{\frac{1}{3}}}{p_{*i-1 \,i}}\,.
\end{equation}
The effects of the individual cooling phases sum up weighted by
 $({C}_{0\, i})^{\frac{1}{3}}$. Thus, whenever the radio plasma is most
 extended, cooling is inefficient.

It remains to provide the parameters describing the different
phases. Suppose we want to describe a phase $i$ where the expansion or
compression is described by $b_i$, and two of the following three
quantities are given: $\tau_i$, the time scale of expansion, ${C}_{i-1\,
i}$ the compression ratio during the phase, and $\Delta t_i$ the
duration of the phase. Theses quantities are related via
\begin{equation}
\label{eq:Retc}
{C}_{i-1\, i} = (1+\Delta t_i/\tau_i)^{-b_i}\,.
\end{equation}
We get
\begin{equation}
\label{eq:p*i-1i}
p_{*i-1\, i} = \frac{{C}^{\frac{1}{3}}}{a_{0} \, t_i \, \left(
\frac{{C}^{5/3} - {C}^{1/b}}{1- 5\,b /3}
\,u_{B\,i-1} + \frac{{C}^{1/3} - {C}^{1/b}}{1- b
/3}\, u_{C} \right)}\,,
\end{equation}
with ${C} = {C}_{i-1\, i}$ and $t_i = \tau_i + \Delta t_i$ for
brevity. The magnetic energy density at the beginning of phase $i$ is
that of the end of phase phase $i-1$:
\begin{equation}
u_{B\,i-1} = u_{B\,0} \, ({C}_{0\,i-1})^{4/3}\,.
\end{equation}
The resulting electron spectrum from an initial power-law distribution
is
\begin{equation}
\label{eq:spec3}
f_i(p) = \tilde{f}_{0} \, {C}_{0\,i}^{\frac{\alpha_{\rm e} +2}{3}}\,
p^{-\alpha_{\rm e}}\, \left( 1 - p/p_{*0\,i} \right)^{\alpha_{\rm e} -2}\,,
\end{equation}
for $p_{{\rm min}\,i} = p_i(p_{\rm min \,0}) < p < p_{{\rm max}\,i} =
p_i(p_{\rm max \,0})$ and $f_i(p) = 0 $ otherwise.
The synchrotron emission at a given frequency $\nu$ is
\begin{equation}
L_{\nu \,i} = c_3 \, B_i\,V_i\, \int_{p_{{\rm min}\,i}}^{p_{{\rm max} \,i}}
\!\!\!\! \!\!\!\!  \!\!\!\! d p\, f_i (p)\, \tilde{F}(\nu/ \nu_{i}(p)),
\end{equation}
where $c_3 = \sqrt{3}\, e^3/(4\, \pi\, m_{\rm e}\, c^2)$ and the
characteristic frequency is $\nu_{i}(p) = 3\, e\, B_i\, p^2
/(4\,\pi\, m_{\rm e}\, c)$. The dimensionless spectral emissivity of a
mono-energetic isotropic electron distribution in isotropically
oriented magnetic fields $\tilde{F}(x)$ can be approximated
\cite{1999AA...344..409E}:
\begin{equation}
\tilde{F}(x) \approx \frac{2^{2/3} \, (\pi/3)^{3/2}}{ \Gamma(11/6) }
 x^{1/3} \exp \left( -\frac{11}{8} x^{7/8}\right)\,.
\end{equation}
In reality, after shock passage an originally isotropic ensemble of
field lines gets partially aligned with the shock plane.  This is also
true for the unshocked radio plasma, since its morphology gets
significantly flattened during compression (see phase 3 in
Sec. \ref{sec:phases}). This would produce a radio polarization and a
luminosity which depends on the viewing angle
\cite{1998AA...332..395E}. As long as we are only calculating the
total luminosity of the radio cocoon/relic, this can be ignored. But,
in case one wants to know the expected flux, one has to correct for
the anisotropic emission pattern of the cluster radio
relics. Fortunately, the degree of radio polarization can be used to
determine the viewing angle with respect to the shock plane
\cite{1980MNRAS.193..439L,1998AA...332..395E}.

The upper cutoff in the electron distribution at $p_{*0\,n}$ produces
a cutoff in the synchrotron spectrum near $\nu_{* n} = \nu_{\rm
c}(p_{*0\,n})$. But since $\tilde{F}(x)$ has a broad maximum even a
sharp cutoff in the electron spectrum gives a soft cutoff in the
radio, with significant flux above $\nu_{* n}$.

\begin{table*}[t]
\begin{tabular}{|l|llll|llllrlll|}
\hline
			&  
$\Delta t_i$ 		&  
$\tau_i$    		&  
$C_{i-1\,i}$ 		&  
$b_i$     		&  
${C}_{0\,i}$ 		&  
$V_n$ 			&  
$B_i$        		&  
$u_{B\,i}$  		&  
$p_{*0\, i}$  		&  
$\nu_{* \,i}$		&  
$L_{\rm 0.1\,GHz}$ 	&  
$L_{\rm 1\,GHz}$ 	\\  
                        &  
Gyr			&  
Gyr			&  
   			&  
   			&  
   			&  
${\rm Mpc^3}$		&  
$\mu$G 			&  
${\rm eV\,cm^{-3}}$	&  
   			&  
GHz			&  
\multicolumn{2}{c|}{$10^{31}\,{\rm erg\,s^{-1}\,Hz^{-1}}$}\\  
\hline
Scenario A &&&&&&&&&&&&\\
 Phase 0			 &  
$0$			 &  
$0.015$			 &  
$1.$			 &  
$1.8$			 &  
$1.$			 &  
$0.0059$			 &  
$12.$			 &  
$3.6$			 &  
$\infty$			 &  
$\infty$			 &  
$530.$			 &  
$93.$			 \\  
 Phase 1			 &  
$0.0054$			 &  
$0.01$			 &  
$0.595$			 &  
$1.2$			 &  
$0.595$			 &  
$0.0099$			 &  
$8.5$			 &  
$1.8$			 &  
$36700.$			 &  
$48.$			 &  
$210.$			 &  
$33.$			 \\  
 Phase 2			 &  
$0.1$			 &  
$\infty$			 &  
$1.$			 &  
$0$			 &  
$0.595$			 &  
$0.0099$			 &  
$8.5$			 &  
$1.8$			 &  
$2730.$			 &  
$0.27$			 &  
$86.$			 &  
$0.22$			 \\  
 Phase 3			 &  
$0.069$			 &  
$-0.11$			 &  
$6.45$			 &  
$2.$			 &  
$3.83$			 &  
$0.0015$			 &  
$29.$			 &  
$22.$			 &  
$1220.$			 &  
$0.19$			 &  
$1500.$			 &  
$0.94$			 \\  
 Phase 4			 &  
$0.02$			 &  
$\infty$			 &  
$1.$			 &  
$0$			 &  
$3.83$			 &  
$0.0015$			 &  
$29.$			 &  
$22.$			 &  
$651.$			 &  
$0.052$			 &  
$240.$			 &  
$0$			 \\  
\hline
Scenario B &&&&&&&&&&&&\\
 Phase 0			 &  
$0$			 &  
$0.015$			 &  
$1.$			 &  
$1.8$			 &  
$1.$			 &  
$0.12$			 &  
$2.7$			 &  
$0.18$			 &  
$\infty$			 &  
$\infty$			 &  
$38.$			 &  
$6.7$			 \\  
 Phase 1			 &  
$0.17$			 &  
$0.01$			 &  
$0.0316$			 &  
$1.2$			 &  
$0.0316$			 &  
$3.7$			 &  
$0.27$			 &  
$0.0018$			 &  
$8810.$			 &  
$0.088$			 &  
$0.0096$			 &  
$0$			 \\  
 Phase 2			 &  
$1.$			 &  
$\infty$			 &  
$1.$			 &  
$0$			 &  
$0.0316$			 &  
$3.7$			 &  
$0.27$			 &  
$0.0018$			 &  
$1830.$			 &  
$0.0038$			 &  
$0$			 &  
$0$			 \\  
 Phase 3			 &  
$0.55$			 &  
$-0.67$			 &  
$31.6$			 &  
$2.$			 &  
$1.$			 &  
$0.12$			 &  
$2.7$			 &  
$0.18$			 &  
$3270.$			 &  
$0.12$			 &  
$7.4$			 &  
$0.00035$			 \\  
 Phase 4			 &  
$0.3$			 &  
$\infty$			 &  
$1.$			 &  
$0$			 &  
$1.$			 &  
$0.12$			 &  
$2.7$			 &  
$0.18$			 &  
$1910.$			 &  
$0.041$			 &  
$0.98$			 &  
$0$			 \\  
\hline
Scenario C &&&&&&&&&&&&\\
 Phase 0			 &  
$0$			 &  
$0.015$			 &  
$1.$			 &  
$1.8$			 &  
$1.$			 &  
$0.0059$			 &  
$1.2$			 &  
$0.036$			 &  
$\infty$			 &  
$\infty$			 &  
$0.3$			 &  
$0.053$			 \\  
 Phase 1			 &  
$0.032$			 &  
$0.01$			 &  
$0.178$			 &  
$1.2$			 &  
$0.178$			 &  
$0.033$			 &  
$0.38$			 &  
$0.0036$			 &  
$54800.$			 &  
$4.8$			 &  
$0.014$			 &  
$0.00099$			 \\  
 Phase 2			 &  
$0.1$			 &  
$\infty$			 &  
$1.$			 &  
$0$			 &  
$0.178$			 &  
$0.033$			 &  
$0.38$			 &  
$0.0036$			 &  
$16200.$			 &  
$0.42$			 &  
$0.0078$			 &  
$0.000028$			 \\  
 Phase 3			 &  
$0.13$			 &  
$-0.17$			 &  
$15.9$			 &  
$2.$			 &  
$2.83$			 &  
$0.0021$			 &  
$2.4$			 &  
$0.14$			 &  
$16500.$			 &  
$2.7$			 &  
$1.4$			 &  
$0.12$			 \\  
 Phase 4			 &  
$0.1$			 &  
$\infty$			 &  
$1.$			 &  
$0$			 &  
$2.83$			 &  
$0.0021$			 &  
$2.4$			 &  
$0.14$			 &  
$7860.$			 &  
$0.62$			 &  
$1.$			 &  
$0.02$			 \\  
\hline
\end{tabular}
\caption[]{\label{table} Parameter of the three szenarios. 
The rows labled with 'Phase 0' give the initial parameters of the
radio plasma at the end of the activity of the radio galaxy. The rows
below ('Phase 1-4') give the subsequent evolution, due to aging and
expansion or compression described by the left block of parameters.
Details can be found in the text. The luminosity units allow an easy
conversion to the observable flux: for a source in $100\,{\rm Mpc}$
distance the flux of $1\,{\rm Jansky}$ corresponds to a luminosity of
$1.2\cdot 10^{31}\,{\rm erg\,s^{-1}\,Hz^{-1}}$.}
\end{table*}

\section{The Model\label{sec:phases}}

Between the release from a radio galaxy and the reappearance as a
cluster radio relic the radio plasmon undergoes several different
phases of expansion and contraction:

{\bf Phase 0: Injection.} The radio galaxy is active and a large
expanding volume is being filled with radio plasma. The expansion of
this cocoon is likely to be supersonic (with respect to the outer
medium) and therefore $b_0 = 9/5$, if we assume that there is no gas
density gradient in the vicinity of the radio galaxy
\cite{1997MNRAS.286..215K}. We assume that the injection occurred with
a time constant of $\tau = 0.15\,{\rm Gyr}$, which is the typical age
of a radio source at the end of nuclear activity \cite[for typical
ages]{1987MNRAS.225....1A}.  The particle population is kept close to
a power-law distribution by injection of fresh electrons. We assume a
spectral index of $\alpha_{\rm e} = 2.5$ and an upper cutoff at $p_{\rm
max\, 0} = 10^5$. The momentum cut-off produces a radio cut-off above
$42\,{\rm GHz}\,(B/\mu{\rm G})$, but we note that our results are
not very sensitive to the choice of this parameter.  More realistic
electron spectra at the end of phase 0 could be constructed by
superimposing the spectra of the electron populations of different 
ages, as in the models of Kaiser et al. \cite*{1997MNRAS.292..723K}. 
But, for the
present purpose of demonstrating that the fossil radio plasma can be
revived by compression, our simplified treatment should be
sufficiently illustrative.

{\bf Phase 1: Expansion.} After the central engine of the radio galaxy
became inactive, the radio cocoon might still be strongly
over-pressured compared to its gaseous environment
\cite{1989ApJ...345L..21B}.  If this is the case a Sedov-like
expansion phase exists with $b_1 = 6/5$.  Momentum conservation of the
expanding shell around the cocoon forces the expansion rates at the
end of phase 0 and beginning of phase 1 to be the same, leading to
$\tau_1 = b_1/b_0 \, \tau_0 = 2/3\, \tau_0$.  The expansion will
significantly deviate from $b_1= 6/5$ at the moment when the
internal pressure drops to a value comparable to the environmental
pressure.  We simplify this behavior by assuming that the expansion is
Sedov-like until pressure equilibrium is reached. The radio cocoon
probably becomes undetectable during this phase, and therefore
becomes a fossil radio cocoon or a so called radio ghost.

{\bf Phase 2: Lurking.} Pressure equilibrium with the environment is reached,
and the volume of the radio plasma remains more or less constant. Thus
${C}_{1\, 2} \approx 1$ and $b \approx 0$. Taking the appropriate limit
$\tau_2 \rightarrow \infty$ of Eq. \ref{eq:p*i-1i} gives the well known result
\begin{equation}
\label{eq:p*12}
p_{*1\, 2} = \left[a_{0} \, \left( \,u_{B\,2} + u_{C}\right) \, \Delta
t_2 \right]^{-1}\,.
\end{equation}
Due to the previous adiabatic energy losses of the electrons, they
reside at low energies during phase 2. Their radiation losses, which
strongly depend on the particle energies, are therefore strongly
diminished.  Additionally, the synchrotron losses are further reduced
due to the weaker magnetic field during the expanded state of the
radio cocoon. The adiabatic losses are reversible, and will be
reversed during the subsequent compression phase, whereas the
radiative losses are irreversible. Since the latter are suppressed
during this phase, the radio ghost state can be called the {\it
energy saving mode} of a radio cocoon.

{\bf Phase 3: Flashing.} The fossil radio plasma gets dragged into a shock
wave of cosmic large-scale structure formation, e.g. at the boundary
of a cluster of galaxies or in a galaxy cluster merger event. While its
thermal environment gets shocked, the radio plasma itself is only
compressed adiabatically, due to the much higher internal sound speed.
The electron population and the magnetic field gain energy
adiabatically, leading to a steep enhancement of the synchrotron 
emissivity.

This phase has a duration of the order of $\Delta t_3 = V_2^{1/3}
/v_{\rm shock\, 2} \approx $ $ {\rm 0.1 ... 1\, Mpc/ (300 ... 3000 \,
km/s)} \approx $ $ {\rm 30\, Myr ... 3\, Gyr}$. The pre-shock flow
velocity in the shock frame, $v_{\rm shock\, 2}$, is related to the
pre-shock sound speed $c_{\rm s\,2}= \sqrt{\gamma\,P_2/(n_{\rm e}\,(m_{\rm p} +
m_{\rm e})/2)}$ via
\begin{equation}
v_{\rm shock \,2}^2 = \frac{c_{\rm s\, 2}^2}{2 \gamma} \,\left(\gamma
-1 + (\gamma+1) \frac{P_3}{P_2}\right)
\end{equation} 
\cite{landau.hydro}, where $\gamma = 5/3$ is the adiabatic index of
the thermal gas.  The compression factor of the relativistic plasma is
high, and can be calculated from the assumed pressure jump $P_3/P_2$
of the surrounding thermal gas,
assuming pressure equilibrium before 
and after the shock passage:
\begin{equation} 
{C}_{2\, 3} = (P_3/P_2)^\frac{3}{4}\,.
\end{equation}

In order to guess the parameters $b_3$ and $\tau_3$ a rough picture of
the compression process is required. The sound speed within the radio
plasma should be much higher than even in the post-shock thermal
environment, (it could be up to $c/\sqrt{3}$, if the plasma is fully
relativistic), so that an instantaneous response to environmental
changes can be assumed. During the shock passage, the cocoon is exposed
to the high thermal pressure of the post-shock gas on its down- and to
the ram-pressure of the pre-shock gas on its up-stream
side.  But, on the remaining surface the cocoon is only subject to the
(much lower) up-stream gas pressure. The relativistic plasma will
therefore start to expand orthogonal to the gas flow, producing a
flattened pancake-like morphology. In order be able to expel the
ambient gas sideways additional internal pressure comparable to the
ram pressure of the expelled material is needed. This pressure is
produced by compression.  The process of flattening stops when the ram
pressure of the swept-up material at the expanding edges of the
pancake is of the order of the ram pressure of the incoming flow. This
implies that the ratio of the diameter to the thickness of the
expanded cocoon is roughly 4 for a strong shock.

The compression is slow in the beginning, and rapid towards the end
when the cocoon is significantly flattened. We mimic this by setting
$b_3 = 2$ and a negative $\tau_3$, according to Eq. \ref{eq:Retc}. We
favor this over the more intuitive choice of a positive $\tau_3$ and
large negative $b_3$, since it describes the process of compression
more realistically.

{\bf Phase 4: Fading.} The radio plasma is in pressure equilibrium
with the post-shock medium, which should provide roughly a uniform
environment: $b_4 = 0$. The radio emission of the relic now fades away
due to the heavy radiation losses.

\section{The Scenarios\label{sec:scenarios}}
We follow the evolution of a cocoon of radio plasma for three
different scenarios (A, B, and C) and calculate the resulting radio
emission.  These scenarios are envisioned to illustrate plausible
situations, rather than try to reproduce precise parameters for
some known cluster radio relics.  Scenario A and B are chosen to
represent extreme cases: In A, the relic is located at the
center of a galaxy cluster, while in B, its location is near the
cluster boundary, i.e., in the proximity of the accretion shock
wave. In both scenarios the duration of phase 2 was chosen to be so
long that the shocked radio plasma could be barely observed as a weak
ultra-steep spectrum source. Since these scenarios are rather extreme,
we demonstrate with scenario C that a shorter phase 2 can result in a
moderately steepened spectrum of the cluster radio relic.

For the cases A and B, we assume the initial cocoon (at the end of phase 0)
to contain magnetic fields, relativistic electrons and protons with
energies of $E_{\rm B/e/p} = 10^{60} {\rm erg}$ each, and $E_{\rm
B/e/p} = 10^{58} {\rm erg}$ in case C.  The three components produce a
relativistic isotropic pressure of
\begin{equation}
\label{eq:Pcocoon}
P_{\rm cocoon\, 0} = \frac{E_{\rm e} + E_{\rm p} + E_B}{3\,V_0}\,.
\end{equation}

We assume a spectral index of $\alpha_{\rm e} = 2.5$ and a rather high
cutoff in the electron spectrum at $p_{\rm max \,0} = 10^5$, which has no
significant influence on the conclusions. The lower cutoff is $p_{\rm
min\,0} = 10$ in scenarios A and B, and $p_{\rm min\,0} = 100$ in
C. The lower cutoffs only affect the normalizations of the radio
fluxes, not the spectral shapes.

\begin{figure}[t]
\begin{center}
\psfig{figure=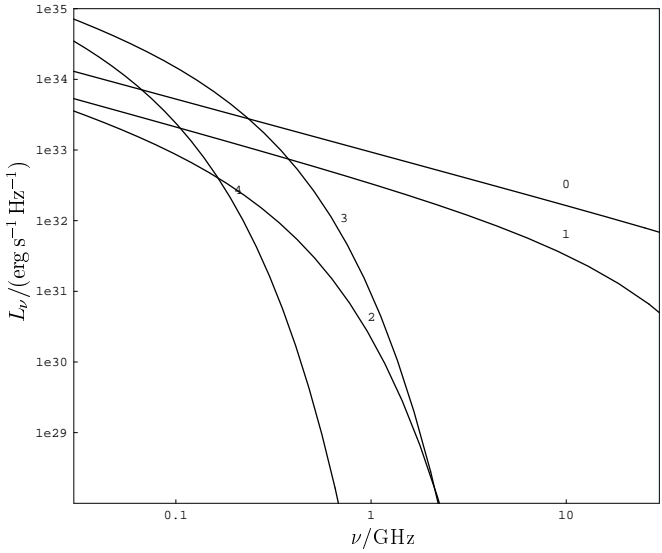,width=0.45 \textwidth,angle=0}
\end{center}
\caption[]{\label{fig:syncA} Radio spectrum of the radio cocoon in
scenario A at the end of phases 0-4.}
\end{figure}

\begin{figure}[t]
\begin{center}
\psfig{figure=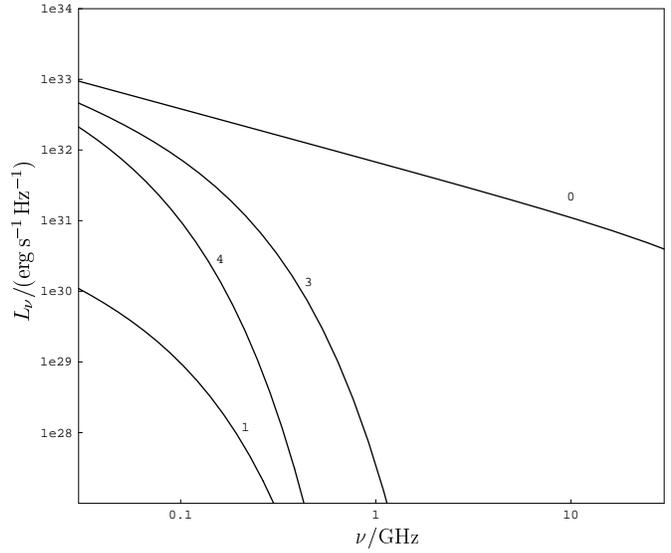,width=0.45 \textwidth,angle=0}
\end{center}
\caption[]{\label{fig:syncB} Radio spectrum of the radio cocoon in
scenario B at the end of phases 0-4. The luminosity at the end of
phase 2 is too small in order to be displayed in this figure.}  
\end{figure}

\begin{figure}[t]
\begin{center}
\psfig{figure=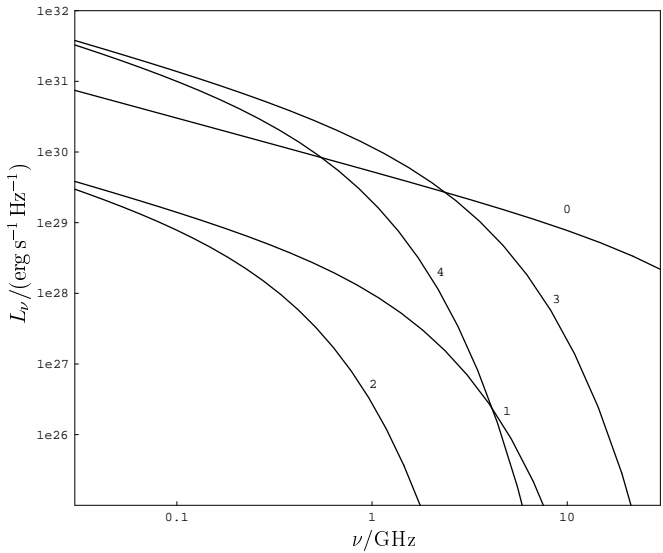,width=0.45 \textwidth,angle=0}
\end{center}
\caption[]{\label{fig:syncC} Radio spectrum of the radio cocoon in
scenario C at the end of phases 0-4.}
\end{figure}

The different scenarios are described below in detail. The parameters
for the different scenarios are listed in Tab. \ref{table} and the 
resulting radio spectra at the end of the different phases are shown
in Fig \ref{fig:syncA}, \ref{fig:syncB}, and \ref{fig:syncC}.

\noindent
{\bf Scenario A: The Cocoon at the Cluster Center.}  The pre-merger
cluster is assumed to have an electron density of $n_{\rm e} = 0.3\cdot
10^{-3} \,{\rm cm^{-3}}$ and a temperature of $kT = 3\,{\rm keV}$ at
the location of the radio galaxy. Due to the high environmental
pressure, we assume the internal pressure of the cocoon to be only
twice the external pressure, and use Eq. \ref{eq:Pcocoon} to obtain
the initial volume $V_0$. Due to the strong synchrotron energy losses
in this scenario, phase 2 can not last much longer than $\Delta t_2 =
0.1 \,{\rm Gyr}$, otherwise the revived fossil radio cocoon would
not emit within the observable radio frequency range.  We assume that
the shock wave of a cluster merger event increases the internal
pressure by $P_3/P_2 = 12$ during the phase 3. This corresponds to a
moderate shock with shock compression factor of 2.8, whereas strong
non-relativistic shocks can have a compression factor of 4. A moderate
shock is expected, since both merging clusters are expected to have
temperatures of several keV and therefore sound velocities comparable
to the merger velocity.

As can be seen in Fig. \ref{fig:syncA} the compression caused by the
merger shock wave gives rise to a burst of low frequency emission, but
practically no high frequency emission. This is due to the rapid
decay of the upper end of the electron spectrum during phase 3,
which essentially wipes out the adiabatic energy gains of these
electrons. The source decays on a time-scale of a few tens of Myr,
mostly due to the heavy synchrotron losses. If the radio cocoon is
located in a more peripheral region of the cluster, where the density,
the pressure and therefore the magnetic field strength inside the
cocoon is much lower, these losses are also much milder. 
This lengthens the time scale over which the radiatively cooling
synchrotron plasma can still be revived by the next passing shock, and
thus rendered radio detectable.  We, therefore, expect the radio relic
phenomena to be found preferentially at larger cluster radii, and less
often near the cluster center (although projection can help some
relics to appear near the cluster core). The best environment to find
cluster radio relics is, therefore, near the edges of the clusters.

\noindent
{\bf Scenario B: The Cocoon at the Cluster Boundary.} The radio cocoon
is assumed here to be born outside the cluster, in an environment of a
dense galaxy filament, or a group of galaxies, say, with $n_{\rm e} =
0.3\cdot 10^{-5} \, {\rm cm^{-3}}$ and $kT = 0.3\, {\rm keV}$.  The
freshly injected radio plasma might be over-pressured by a factor of
100, leading to a short expansion phase. After this, the electrons
within the expanded Mpc sized cocoon suffer mostly the IC-losses,
allowing revival of the radio plasma even $\Delta t_2 = 1 \, {\rm
Gyr}$ later. This can happen when the cocoon along with the ambient
medium is crossed by the accretion shock of a cluster of galaxies,
which might entail a pressure jump as large as $P_3/P_2 = 100$, in
order to heat the infalling cool gas to the cluster virial temperature
of up to $10\,{\rm keV}$.

\begin{figure}[t]
\begin{center}
\psfig{figure=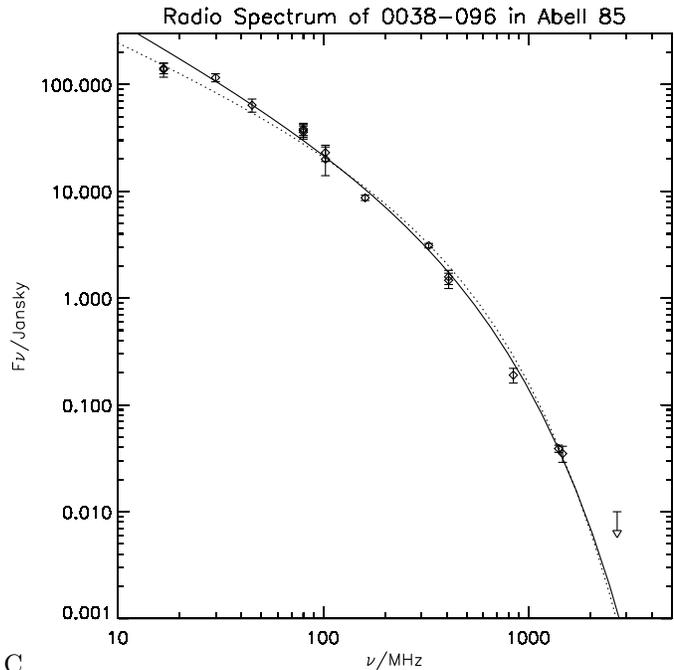,width=0.45 \textwidth,angle=0}
\end{center}
\caption[]{\label{fig:A85} Radio spectrum of the radio relic in A85.
The data was compiled from the literature by Bagchi et
al. \cite*{1998MNRAS.296L..23B}.  The solid line is a radio spectrum
resulting from the canonical form (Eq. \ref{eq:spec3}) with $\alpha_{\rm e}
= 3$ and $p_{*} = 11000\,(B/{\mu{\rm G}})^{-1/2}$, and the dotted line
one with $\alpha_{\rm e} = 2.65$ and $p_{*} = 10000\,(B/{\mu{\rm
G}})^{-1/2}$.}
\end{figure}

Scenario B can explain the steep and bent radio spectrum of the
cluster radio relic 0038-096 in Abell 85. An eye-fit to the radio
spectrum (Fig. \ref{fig:A85}) shows that the maximal electron momentum
in this case is $p_* = 10^4\,\,(B/{\mu{\rm G}})^{-1/2}$. The magnetic
field strength of the cluster relic was estimated from the minimum
energy argument to be $B \approx 1\,\mu$G \cite{1996IAUS..175..333F}
and from the detection of excess X-ray emission at the location of the
relic, which implies a field strength of $B = 0.95 \pm 0.10 \,\mu$G
\cite{1998MNRAS.296L..23B} if this emission refers to the IC scattered
cosmic microwave background photons, otherwise a higher field
strength. Using $B= 1\,\mu$G and $p_* = 10^4$ and assuming a uniform
environment without expansion and compression, an age of $0.2\,{\rm
Gyr}$ would result \cite[and see
Eq. \ref{eq:p*12}]{1994A&A...285...27K}.  But scenario B demonstrates
that the radio plasma can be as old as $2\,{\rm Gyr}$. This resolves
the problem of the apparent cooling time of the electrons being too
short for any nearby galaxy to have ejected the plasma and then moved
to its present location with a typical velocity of a cluster member.
For the long duration of phase 2 the resulting spectrum is fairly
steep in the observable radio range. But this need not to be the case
for a scenario with a shorter fossil phase.

\noindent
{\bf Scenario C: The Smoking Gun.}  In order to substantiate the last
statement, we choose a set of parameters for scenario C which produces
a cluster radio relic with relative flat, nearly unbent radio
spectrum, like the relic 1253+275 at the boundary of the Coma cluster
of galaxies 
\cite[and see Fig. \ref{fig:ngc4789}]{1991A&A...252..528G}.  
There, the probable source of the radio plasma is visible: it is the
galaxy NGC 4789, which is located upstream of the relic, and which
seems to be on an ascending orbit after a cluster core passage
\cite{1998AA...332..395E}. Its narrow angle tailed radio outflow --
{\it the smoke of the gun NGC 4789} -- seems to be bent and dragged by
the infalling matter (falling into Coma's gravitational potential) to
the location of the relic. There, it brightens up and exhibits a steep
but straight spectrum with a slope of $1.18$. For details of the
spectrum and the geometry, see Giovannini et
al. \cite*{1991A&A...252..528G} and En{\ss}lin et
al. \cite*{1998AA...332..395E}.

We follow a blob of the radio plasma, which might have been injected
with an overpressure of a factor 10 into the infalling gas stream with
possibly $n_{\rm e} = 0.3\cdot 10^{-5}\,{\rm cm^{-3}}$ and $kT = 0.6\, {\rm
keV}$. The inflow of the plasma and the compression at the
cluster accretion shock wave might have taken a few hundreds of
Myr. We assume a pressure jump of only $P_3/P_2 = 40$ at the shock
wave, not higher, in order to allow the temperature of the post-shock
gas to stay below the average cluster temperature of 8.2 keV
\cite{1992A&A...259L..31B}. As can be seen in Fig. \ref{fig:syncC}, the
radio spectrum below 1 GHz stays practically unbent for a couple of 
tens of Myr after the shock passage.

\section{Discussion\label{sec:discussion}}

We have argued here that adiabatic compression in cluster shocks can
revive fossil radio plasma to radio detection, even up to 2 Gyr after
its release from the parent radio galaxy. The computed radio spectra
provide a good match to the observed spectra of cluster radio
relics. Hence, this is a promising model for the regions of diffuse
radio emission found in clusters of galaxies. Below, we
summarize some merits and potential shortcoming of this model:\\

\noindent
{\bf Pros:}
\begin{itemize}
\item The observed connection of cluster radio relics to shock waves
arises naturally in this model.
\item The presence of the radio galaxy NGC 4789 and the morphological
connection of its radio tails to the relic 1253+275 seems to be the
{\it smoking gun} of cluster radio relic formation by compression of
fossil radio plasma.
\item The expected very high sound speed within radio plasma should
virtually forbid shock waves of the ambient medium to penetrate into
radio cocoons. This renders adiabatic compression as a more plausible
means of reviving the fossil plasma.
\item Cluster radio relics are rare. Not only a shock wave and fossil
radio plasma have to be present in this model, in order to produce a
cluster radio relic, but also the fossil radio plasma can not grow too
old (of the order of 0.1 Gyr in the center of clusters and 1 Gyr at
their boundaries). Otherwise the high energy end of the electron
population would suffer too severe a depletion for the adiabatic gains
during the compression to be able to shift them to radio emitting
energies. Since fossil radio plasma should be very common
\cite{ensslin2000a}, this helps to explain the observed rarity of
cluster radio relics.
\item The observed tendency of relics to appear at peripheral
positions in clusters follows in this model from the much shorter
lifetimes of the radio emission and, consequently, much shorter ages
which the revive-able fossil radio cocoons can have in the center
of clusters, owing to the stronger synchrotron losses occurring there.
\end{itemize}

\noindent
{\bf Contras:}
\begin{itemize}
\item If the radio plasma has a high mass load, due to undetectable
cool gas, it should get shocked, ensuing Fermi-I shock acceleration
{to produce the observed radio emitting electron population
\cite{1998AA...332..395E}}.  Still, the adiabatic gains of the
relativistic particles are inescapable, although restricted in this
case due to the limited compression factor of a shock wave. 
\item {In the case when the fossil radio plasma is supported by a
very hot, non-relativistic gas component, both adiabatic and shock
 acceleration are ineffective. The first due to the lower compressibility 
of a non-relativistic equation of state, while the second
due to the inability of environmental shock waves to penetrate into
the plasma. 
The fact that we do observe cluster radio relics tells us that this 
situation does not necessarily arise (or, alternatively, cluster radio 
relics are not related to radio ghosts).}
\item
The giant cluster radio relic 2006-56 in Abell 3667 shows a rim of
flat radio emission (spectral index around 0.5) at its sharp outer
edge \cite{1997MNRAS.290..577R}. This could be understood in a Fermi-I
shock acceleration model due to the different ages of the electrons
found at different distances from the energizing shock wave
\cite{1998AA...332..395E}. There are two possibilities to explain the
observation of a flat spectrum rim within the adiabatic compression
model: (a) there is a high mass load in the radio plasma, allowing the
shock wave to successively compress the plasma, which should also
produce a spatially different electron population, and (b) the
apparent rim of a flat spectral index is an artifact arising from
an unmatched coverage of the Fourier space in the radio
interferometric observations at the two frequencies used.
\end{itemize}
We conclude that the model presented here provides a fairly natural
and promising explanation for the phenomenon of cluster radio
relics. This model predicts the existence of a population of diffuse,
ultra-steep, very low frequency radio sources inside and possibly also
outside of clusters of galaxies, due to the age dependence of the
upper frequency cutoff of radio emission arising from the revivable
fossil radio plasma. The ongoing rapid improvements in the
sensitivities at low radio frequencies (Giant Meterwave Radio
Telescope: (GMRT), Low Frequency Array (LOFAR)) may therefore open a
new window on cosmological structure formation by detecting shock
waves marked by such relic radio sources.

\acknowledgements 

We would like to acknowledge Christian Kaiser and Peter Biermann for
discussions and comments on the manuscript. We also thank Joydeep
Bagchi, Vincent Pislar and Gastao Lima Neto for the compiled radio
spectrum of 0038-096 and Gabriele Giovannini, Luigina Feretti, and
Carlo Stanghellini for the radio data of 1253+275.  We thank the
anonymous referee for stimulating comments. TAE thanks for the
friendly hospitality at the {\it National Center for Radio
Astrophysics (NCRA)} in Pune, where this work was initiated. He
further thanks the DFG and the SOC of the IAU 199 conference for
additional travel support.

\bibliography{tae}
\bibliographystyle{aabib99}
\end{document}